\documentstyle[aps,epsfig,floats]{revtex}
\def\lsi{\raise0.3ex\hbox{$<$\kern-0.75em\raise-1.1ex\hbox{$\sim$}}}
\def\gsi{\raise0.3ex\hbox{$>$\kern-0.75em\raise-1.1ex\hbox{$\sim$}}}

\begin{document}
\twocolumn[\hsize\textwidth\columnwidth\hsize\csname
@twocolumnfalse\endcsname

\title{Ultra High Energy Cosmic Rays from Compact Sources}
\author{
Z.~Fodor and S.D.~Katz}
\address{Institute for Theoretical Physics, E\"otv\"os University,\\
P\'azm\'any 1, H-1117 Budapest, Hungary}

\date{\today}
\maketitle

\vspace*{-4.0cm}
\noindent
\hfill \mbox{ITP-Budapest 558}
\vspace*{3.8cm}

\begin{abstract} \noindent
The clustering of
ultra high energy (above $10^{20}$~eV) cosmic rays 
(UHECR)
suggests that they might be emitted by compact sources. 
Statistical analysis of Dubovsky et al.  
(Phys. Rev. Lett. 85 (2000) 1154)
estimated the source density.
We extend their analysis to 
give also the confidence intervals for the source density using
a.) no assumptions on the relationship between clustered and 
unclustered events; b.) nontrivial distributions for the source 
luminosities and energies; c.) the energy dependence of the propagation.
We also determine the probability that a proton created at a distance $r$ with
energy $E$  arrives at earth above a threshold $E_c$. Using this function
one can determine the observed spectrum just by one numerical integration
for any injection spectrum.
The observed 14 UHECR events above $10^{20}$~eV with one doublet gives 
for the source densities $180_{-165}^{+2730}\cdot 10^{-3}$~Mpc$^{-3}$ 
(on the 68\% confidence level). We present detailed results 
for future experiments with larger UHECR statistics.

\end{abstract}

\vspace*{0.2cm}
\pacs{PACS numbers: 96.40.Pq, 98.70.S}
\vskip1.5pc]

\section{INTRODUCTION}

The interaction of protons with photons of the cosmic microwave 
background predicts a sharp drop in the cosmic ray flux above
the GZK cutoff around $5\cdot 10^{19}$~eV \cite{GZK66}. The 
available data shows no such drop. About 20 events above $10^{20}$~eV
were observed by a number of experiments such as AGASA 
\cite{AGASA}, Fly's Eye \cite{FLY}, Haverah Park \cite{HAVERAH},
Yakutsk \cite{YAKUTSK} and HiRes \cite{HIRES}. Since above
the GZK energy the attenuation length of particles is a few tens
of megaparsecs \cite{YT93} if an ultra high energy cosmic ray (UHECR) is
observed on earth it must be produced in our vicinity (except for UHECR
scenarios based on weakly interacting particles, e.g. neutrinos).

Usually it is assumed that 
at these high energies the galactic and extragalactic magnetic 
fields do not affect the orbit of the cosmic rays, thus they 
should point back to their origin within a few degrees. In 
contrast to the low energy cosmic rays one can use UHECRs
for point-source search astronomy. (For an extragalactic
magnetic field of $\mu$G rather than the usually assumed nG 
there is no directional correlation with the source \cite{FP00}. Note, that
due to the 
Local Supercluster the magnetic field is presumably not less than 10~nG which
results in a Larmor radius of few tens of megaparsecs for protons above 
$10^{20}$~eV.) 
Though there are some peculiar clustered events, which we discuss in 
detail,  the overall distribution of UHECR on the sky is practically isotropic. 
This observation is rather surprising since in principle only a few 
astrophysical sites (e.g. active galactic nuclei \cite{M95} or the extended 
lobes of radio galaxies \cite{RB93}) are capable of  accelerating such 
particles, nevertheless none \cite{ES95} of the UHECR evets came from these 
directions. Hence it is generally believed \cite{B99} that there is no 
conventional astrophysical explanation for the observed UHECR spectrum.

There are several ways to look for
the source inhomogeneity from the energy spectrum and 
spatial directions of UHECRs. One possibility is to 
assume that the source density of UHECRs is proportional to 
the galaxy densities \cite{WFP97}. Another approach is to analyze 
the clustering properties of the unknown sources by some 
correlation length\cite{BW99}.

Clearly, the arrival directions of 
the UHECRs measured by experiments show some peculiar clustering: 
some events are grouped within $ \sim 3^o$, the typical angular 
resolution of an experiment. Above $4\cdot 10^{19}$ eV 92 cosmic ray events 
were detected, including 7 doublets and 2 triplets. 
Above $10^{20}$ eV one doublet out of 14 events were found \cite{Uchi}. 
The chance probability of such a clustering from uniform distribution is 
rather small \cite{Uchi,Hea96}. (Taking the average bin $3^o$
 the probability of generating one doublet out of 14 events is $11\%$.)

The clustered features of the events initiated
an interesting statistical analysis  
assuming compact UHECR sources \cite{DTT00}. The authors found
a large number, $\sim 400$ for the number of 
sources\footnote{approximately $400$ sources
within the GZK sphere results
in one doublet for 14 events. The order of magnitude of this result is
in some sense 
similar to that of a ``high-school'' exercise: what is the minimal 
size of a class for which the probability of having clustered birthdays
--at least two pupils with the same birthdays-- is larger than 50\%. 
In this case the number of ``sources'' is the number of possible 
birthdays $\sim 400$. In order to get the answer 
one should solve $365!/[365^k (365-k)!] < 0.5$, which gives as a minimal size
k=23.}
inside a GZK sphere of 25~Mpc. 
They assumed that \hfill\break
a.) the number of clustered events is much smaller than the total 
number of events (this is a reliable assumption at present
statistics; however, for any number of sources the increase of
statistics, which will happen in the near future, results in more 
clustered events than unclustered),
\hfill\break
b.) all sources have the same luminosity which gives a delta function 
for their distribution (this unphysical choice represents an important 
limit, it gives the smallest source density for a given
number of clustered and unclustered events)\hfill\break
c.) The GZK effect makes distant sources fainter; however, this feature
depends on the injected energy spectrum 
and the attenuation lengths and elasticities of the propagating
particles. In \cite{DTT00} an exponential decay was used with an energy
independent decay length of 25Mpc.

In our approach none of these assumptions are used. In addition
we include spherical astronomy corrections and in particular
give the upper and lower bounds for the source density
at a given confidence level. As we show the most probable value
for the source density is really large; however, the 
statistical significance of this result is rather weak. At
present the small number of UHECR events allows a 95\%
confidence interval for the source density which spreads 
over four orders of magnitude. Since future experiments,
particularly Pierre Auger \cite{B96,G99}, will have a much
higher statistical significance on clustering (the expected 
number of events of $10^{20}$ eV and above is 60 per year 
\cite{BBL00}), we present our results on the density of sources also 
for larger number of UHECRs above $10^{20}$ eV.

In order to avoid the assumptions of \cite{DTT00} a combined analytical 
and Monte-Carlo technique will be presented adopting the 
conventional picture of protons as the ultra high energy cosmic rays.
Our analytical approach of Section \ref{sec_anal}
gives the event clustering 
probabilities for any space, luminosity and energy distribution of 
the sources by using a single additional function $P(r,E;E_c)$, the probability
that a proton created at a distance $r$ with energy $E$ arrives
at earth above the threshold energy $E_c$ \cite{BW99}. 
With our Monte-Carlo technique of Section \ref{sec_monte} we
determine the probability function $P(r,E;E_c)$ for 
a wide range of parameters. Our results 
for the present and future UHECR statistics are presented in Section 
\ref{sec_res} 
We summarize in Section \ref{sec_sum}.  

\section{ANALYTICAL APPROACH} \label{sec_anal}

The key quantity for finding the distribution functions for the
source density is the probability of detecting $k$ events from one randomly
placed source. The number of UHECRs emitted by a source of $\lambda$ luminosity
during a period $T$ follows the Poisson distribution. 
However, not all
emitted UHECRs will be detected. They might loose their energy during
propagation or can simply go to the wrong direction.

For UHECRs the energy loss
is dominated by the pion production in interaction with
the cosmic microwave background radiation. In ref.
\cite{BW99} the probability function $P(r,E,E_c)$ was presented
for three specific threshold energies. This function gives 
the probability that a proton
created at a given distance from earth (r) with some energy (E) is detected
at earth above some energy threshold ($E_c$). The resulting 
probability distribution can be approximated over the energy range 
of interest by a function of the form
\begin{equation}\label{bahcall}
P(r,E,E_c)\approx \exp[-a(E_c)r^2\exp(b(E_c)/E)]
\end{equation}
The appropriate values of $a$ and $b$ for $E_c/(10^{20}{\rm eV})=$1,3, and 
6 are, respectively $a/(10^{-4}{\rm Mpc}^{-2})=$1.4, 9.2 and 11,
$b/(10^{20}eV)=$2.4, 12 and 28.

For the sources we use the
second equatorial coordinate system: ${\bf
x}$ is the position vector of the source characterized by ($r,\delta,\alpha$)
with $\delta$ and $\alpha$ beeing the declination and right ascension,
respectively. The features of the Poisson distribution enforce us to take
into account the fact that the sky is not
isotropically observed. There is a circumpolar cone, in which the sources
can always be seen, with half 
opening angle $\delta'$ ($\delta'$ is the declination of the detector,
for the experiments we study $\delta' \approx 40^o-50^o$). There is also an
invisible region with the same opening angle. Between them there is a region
for which the time fraction of visibility, $\gamma(\delta,\delta')$ is a 
function of the declination of the source. It is straightforward to
determine $\gamma(\delta,\delta')$ for any $\delta$ and $\delta'$:
\begin{equation}
\gamma(\delta,\delta')=\left\{ 
\begin{array}{lll}
0\  &\mbox{if }& -\pi/2< \delta \leq \delta'-\pi/2 \\
1-&\arccos&(\tan\delta'\tan\delta)/\pi \\
 &\mbox{if }& \delta'-\pi/2 <\delta
\leq \pi/2-\delta' \\
1\ &\mbox{if }& \pi/2-\delta' < \delta \leq \pi/2
\end{array}
\right.
\end{equation}
To determine the probability that a particle arriving from random direction
at a random time is detected
we have to multiply $\gamma(\delta,\delta')$ by the cosine of the
zenith angle $\theta$.
In the
following we will use the time average of this function:
\begin{equation}
\eta(\delta,\delta')=\frac{1}{T}\int _0^T{\gamma(\delta,\delta')\cdot
\cos\theta(\delta,\delta',t) dt}
\end{equation}
Since $\delta'$ is constant, in the rest of the paper we do not indicate 
the dependence on it. Neglecting these spherical astronomy 
effects means more than
a factor of two for the prediction of the source density.

The probability of detecting $k$ events from a source at distance 
$r$ with energy $E$ can be obtained by including
$P(r,E,E_c) A\eta(\delta)/(4\pi r^2)$ in the Poisson distribution:
\begin{eqnarray}
p_k({\bf x},E,j)
=\frac{\exp\left[  -P(r,E,E_c)\eta(\delta)j/r^2 \right] }{k!}\times \nonumber\\
\left[ P(r,E,E_c)\eta(\delta)j/r^2\right] ^k, \label{poiss2}
\end{eqnarray}
where we introduced $j=\lambda T A/(4\pi)$ and $A\eta(\delta)/(4\pi r^2)$ 
is the 
probability that an emitted UHECR points to a detector of 
area $A$. 
We denote the space, energy and 
luminosity  distributions of the sources by $\rho({\bf x})$, 
$c(E)$ and $h(j)$, respectively. The probability of detecting $k$
events above the threshold $E_c$ from a single source 
randomly positioned within a sphere of radius $R$ is
\begin{eqnarray}\label{P_k}
P_k=\int_{S_R} dV\; \rho({\bf x}) \int_{E_c}^{\infty} 
dE\; c(E) \int_0^{\infty} dj\; h(j) \times \nonumber \\ 
\frac{\exp\left[ -P(r,E,E_c)\eta(\delta)j/r^2\right] }{k!} \left[
P(r,E,E_c)\eta(\delta)j/r^2 \right] ^k. \end{eqnarray}

Denote the total number of sources within the sphere of 
sufficiently large radius (e.g. several times the GZK radius)
by $N$ and the number of sources that gave $k$ detected events by
$N_k$. Clearly, $N=\sum_0^{\infty}N_i$ and the total number of detected
events is $N_e=\sum_0^{\infty}i N_i$. The probability that for $N$
sources the number of different detected multiplets are $N_k$ is:
\begin{equation}\label{distribution}
P(N,\{N_k\})=N!\prod_{k=0}^{\infty} \frac{1}{N_k!}P_k^{N_k}.
\end{equation}
The value of $P(N,\{N_k\})$ is the most important quantity 
in our analysis of UHECR clustering. For a given set of 
unclustered and clustered events ($N_1$ and 
$N_2,N_3$,...) 
inverting the $P(N,\{N_k\})$ distribution function 
gives the most probable value for the number 
of sources and also the confidence interval for it.
If we want to determine the density of sources we can take the limit
$R \rightarrow \infty, N \rightarrow \infty$, while the density of
sources $S=N/(\frac{4}{3}R^3\pi)$ is constant.

\begin{figure}\begin{center}
\epsfig{file=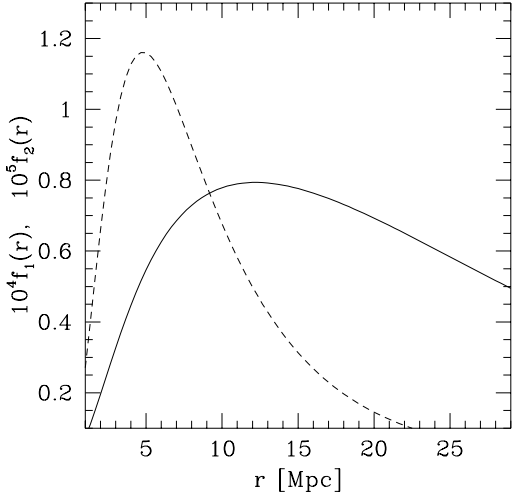,width=7.0cm}
\caption{\label{f_r}
{ \sl The distributions $f_1(r)$ --solid line-- and $f_2(r)$
--dashed line-- of eqn. (\ref{dominance}). The singlet 
and doublet events are dominated by distance scale of 10-15 Mpc
and 3-5 Mpc, respectively.
}}
\end{center}\end{figure}

In order to illustrate the dominant length scale it is instructive 
to study the integrand $f_k(r)$ of the distance integration in 
eqn. (\ref{P_k}) 
\[
P_k=\int_0^R \left(\frac{dr}{R}\right) f_k(r),
\]
\begin{eqnarray}\label{dominance}
f_k(r)=R r^2 \int d\Omega \rho({\bf x}) \int_{E_c}^{\infty} 
dE\; c(E) \int_0^{\infty} dj\; h(j) \times \nonumber \\ 
\frac{\exp \left[ -P(r,E,E_c)\eta(\delta)j/r^2\right] }{k!}
\left[ P(r,E,E_c)\eta(\delta)j/r^2 \right] ^k.
\end{eqnarray}
Fig. \ref{f_r} shows that $f_1(r)$, which leads to singlet events, 
is dominated by the distance scale of 10-15 Mpc, whereas
$f_2(r)$, which gives doublet events, is dominated by the distance
scale of 4-6 Mpc\footnote{
It is interesting that the dominant distance scale for singlet events is
by an order of magnitude smaller than the attenuation length of the protons
at these energies ($l_a\approx 110$~Mpc). This surprising result can be
illustrated using a simple approximation. 
Assuming that the probability of detecting a
particle coming from distance $r$ is proportional to $\exp(-r/l_a)/r^2$,
$P_1$ will be proportional to $\int d\Omega dr r^2\cdot
\exp[-j\exp(-r/l_a)/r^2]\cdot \exp(-r/l_a)/r^2$. For the typical $j$ values
the $r$ integrand has a maximum around 15~Mpc and not at $l_a$.}.
These typical distances partly justify our assumption
of neglecting magnetic fields.
The deflection of singlet events
due to magnetic field does not change the number of multiplets,
thus our conclusions.  The typical distance for higher 
multiplets is quite small, therefore deflection can be
practically neglected. Clearly, the fact that multiplets
are coming from our ``close'' neighbourhood does not mean
that the experiments reflect just the densities of these
distances. The overwhelming number of events are singlets
and they come from much larger distances. Note, that these 
$f_1(r)$ and $f_2(r)$ functions are obtained with our
optimal $j_*$ value (cf. Fig. \ref{ell} and explanation there and in
the corresponding text).
Using the largest possible $j_*$ value allowed by
the 95\% confidence region the dominant distance scales
for $f_1(r)$ and $f_2(r)$ functions turn out to be
30 Mpc and 20 Mpc, respectively.

Note, that $P_k$ and then $P(N,\{N_k\})$ are easily determined by
a well behaving four-dimensional numerical integration
(the $\alpha$ integral can be factorized)
for any $c(E)$, $h(j)$ and $\rho (r)$ distribution functions
. 
In order to illustrate the uncertainties and sensitivities of the
results we used a few different 
choices for these distribution functions. 

For $c(E)$ we studied three possibilities. The most 
straightforward choice is the extrapolation of the `conventional
high energy component' $\propto E^{-2}$. Another possibility is
to use a stronger
fall-off of the spectrum at energies just below the GZK cutoff,
e.g. $\propto E^{-3}$. These choices span the range usually 
considered in the literature and we will study both of them.
The third possibility is to assume that topological defects generate
UHECRs through production of superheavy particles \footnote{Note, that 
these particles are not superheavy DM particles \cite{ELN90}, which are located
most likely in the halo of our galaxy. These superheavy DM particles can
also be 
considered as possible sources of UHECR \cite{BKV97,BS98,BEA} with anisotrpies 
in the arrival direction \cite{PBRS}.}.
According to \cite{BS98}
these superheavy particles decay into quarks and gluons which initiate 
multi-hadron cascades through gluon bremstrahlung. These finally hadronize
to yield jets.
The energy spectrum, first calculated in \cite{Hill} for the
Standard Model and in \cite{BK98} for the Minimal Supersymmetric 
Standard Model, in 
this case can be estimated by the function obtained 
from the HERWIG QCD generator:
\begin{equation}\label{herwig}
c(x)=c_1\frac{\exp[c_2\sqrt{\ln(1/x)}](1-x)^2}{x\sqrt{\ln(1/x)}},
\end{equation}
where $x=E/m_X$ the ratio of the energy and the mass of the decaying
particle. The best fit \cite{BS98} to the 
observed UHECR spectrum gives $m_X\approx 10^{12}$~GeV  for the mass of 
the decaying particle. This corresponds to $c_1\approx 0.0086$ and
$c_2\approx 2.77$. We will use these $c_1$ and $c_2$ values for our
third choice of energy distribution, $c(E)$.

In ref. \cite{DTT00} the authors have shown that for a fixed
set of multiplets the minimal density of sources can be obtained 
by assuming a delta-function distribution for $h(j)$. We 
studied both this limiting case ($h(j)=\delta(j-j_*)$) and a more realistic one
with Schechter's luminosity function \cite{Schechter}:
\begin{equation} \label{spread}
h(j)dj=h\cdot (j/j_*)^{-1.25}\exp(-j/j_*)d(j/j_*). 
\end{equation} 

The space distribution of sources can be given based on some
particular survey of the distribution of nearby galaxies 
\cite{WFP97} or on a correlation length $r_0$ characterizing 
the clustering features of sources \cite{BW99}. For simplicity 
the present analysis deals with a homogeneous distribution of
sources randomly scattered in the universe (Note, that due to the Local 
Supercluster the isotropic distribution is just an approximation.).

\begin{figure}\begin{center}
\epsfig{file=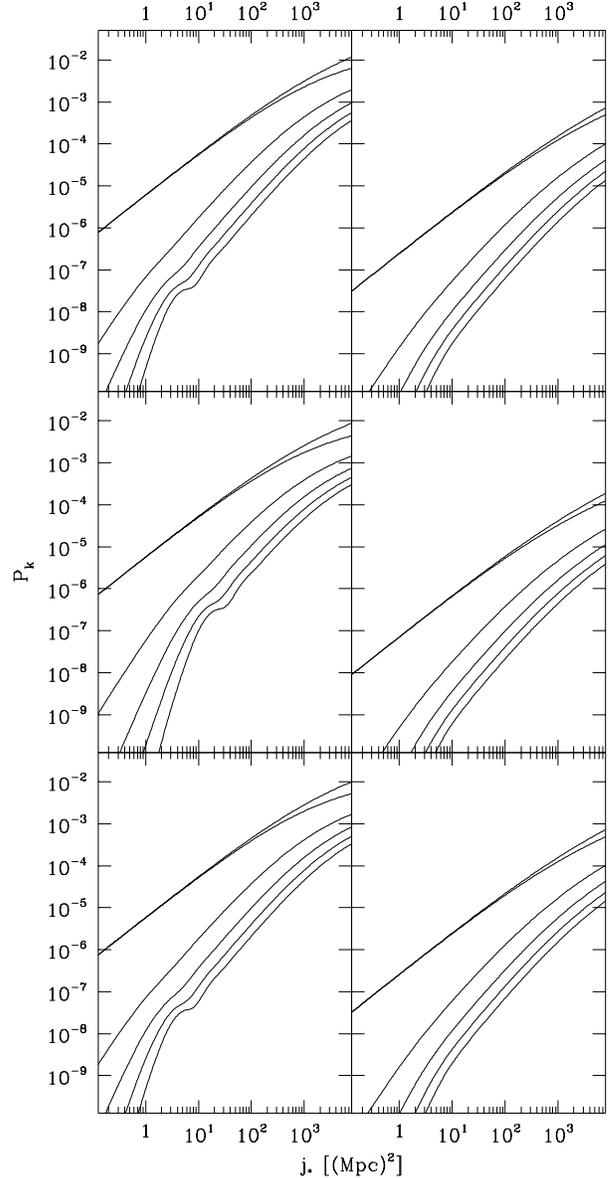,width=8.5cm}
\caption{\label{p_k}
{ \sl 
The individual $P_k(j_*)$ functions for the different $c(E)$ and 
$h(j)$ choices. The column on the left corresponds to the 
Dirac-delta distribution $h(j)=\delta(j-j_*)$, 
whereas the column on the right 
shows the results for Schechter's luminosity distribution.
The first, second and 
third rows correspond to the $c(E)$ functions proportional to
$E^{-2}$, $E^{-3}$ and the superheavy decay mode, respectively
(see text). On each panel the individual lines from top to 
bottom are: $1-P_0$, $P_1$, $P_2$, $P_3$, $P_4$ and $P_5$.  
}}
\end{center}\end{figure}

Fig. \ref{p_k} shows the resulting $P_k(j_*)$ probability functions for the 
different choices of $c(E)$ and $h(j)$. The overall shapes
of them are rather similar; nevertheless, relatively small
differences lead to quite different predictions for the UHECR source
density. The ``shoulders'' of the curves with Dirac-delta
luminosity distributions got smoother for the Schechter's distribution.
The scales on the figures are chosen to cover the 98\% confidence regions
(see section \ref{sec_res} for details). 

Note, that -- assuming that UHECRs point back to their sources -- our
clustering technique discussed above applies to practically 
any models of UHECR (e.g. neutrinos). One only needs a change in the 
$P(r,E,E_c)$ probability distribution function (e.g. neutrinos penetrate the 
microwave background uninhibited) and use the $h(j)$ and $c(E)$ 
distribution function of the specific model. 

\section{MONTE-CARLO STUDY OF THE PROPAGATION} \label{sec_monte}

Our Monte-Carlo model of UHECR studies the propagation
of UHECR.
The analysis of
\cite{DMS98} showed that both AGASA and Fly's 
Eye data demonstrated a change of composition, a shift from heavy 
--iron-- at $10^{17}$~eV to light --proton-- at $10^{19}$~eV
Thus, the chemical composition of UHECRs is most likely
to be dominated by protons. In our analysis we use exclusively protons 
as UHECR particles.
(for suggestions 
that air showers above the GZK cutoff are induced by neutrinos
see \cite{DKDM00}.) 

Using the pion production as the dominant effect of energy loss for
protons at energies $>10^{19}$~eV ref. \cite{BW99} calculated 
$P(r,E,E_c)$, the probability that a proton
created at a given distance (r) with some energy (E) is detected
at earth above some energy threshold ($E_c$). For three
threshold energies the authors of \cite{BW99} gave an approximate 
formula, which we used in the previous section.

In our Monte-Carlo
approach we determined the propagation of UHECR on an event by event 
basis. Since the inelasticity of Bethe-Heitler
pair production is rather small
($\approx 10^{-3}$) we used a continuous energy loss approximation for
this process. The inelasticity of pion-photoproduction is much higher
($\approx 0.2 -0.5$) in the energy range of interest, thus there are only a
few tens of such interactions during the propagation. Due to the Poisson
statistics of the number of interactions and the spread of the
inelasticity, we will see a spread in the energy spectrum even if the
injected spectrum is mono-energetic.

In our simulation protons are propagated in small steps
($10$~kpc), and after each step the energy losses due to pair
production, pion production and the adiabatic expansion are calculated.
During the simulation we keep track of the current energy of the proton
and its total displacement. This one avoids performing new
simulations for different initial energies and distances. The
propagation is completed when the energy of the proton goes below a
given cutoff.
For the proton interaction lengths and inelasticities
we used the values of \cite{BS00,AGNM99}. The deflection 
due to magnetic field is not taken into account, because 
it is small for our typical distances illustrated in Fig. 
\ref{f_r}. This fact justifies our assumption that UHECRs point back to
their sources (for a recent Monte-Carlo analysis on deflection
see e.g. \cite{SEMPR00}).

\begin{figure}\begin{center}
\epsfig{file=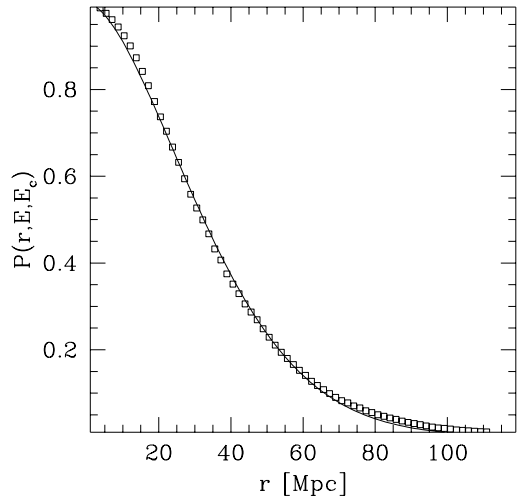,width=7.0cm}
\caption{\label{fit}
{\sl The direct Monte-Carlo points and the fitted
function $P(r,E,E_c)=\exp\left[ -a\cdot(r/\ {\rm 1Mpc})^b\right]$ for
$E_c=10^{20}$~eV and $E=2\cdot 10^{20}$~eV. The fitted curve
corresponds to $a=0.0019$ and $b=1.695$.
}}
\end{center}\end{figure}

\begin{figure}\begin{center}
\epsfig{file=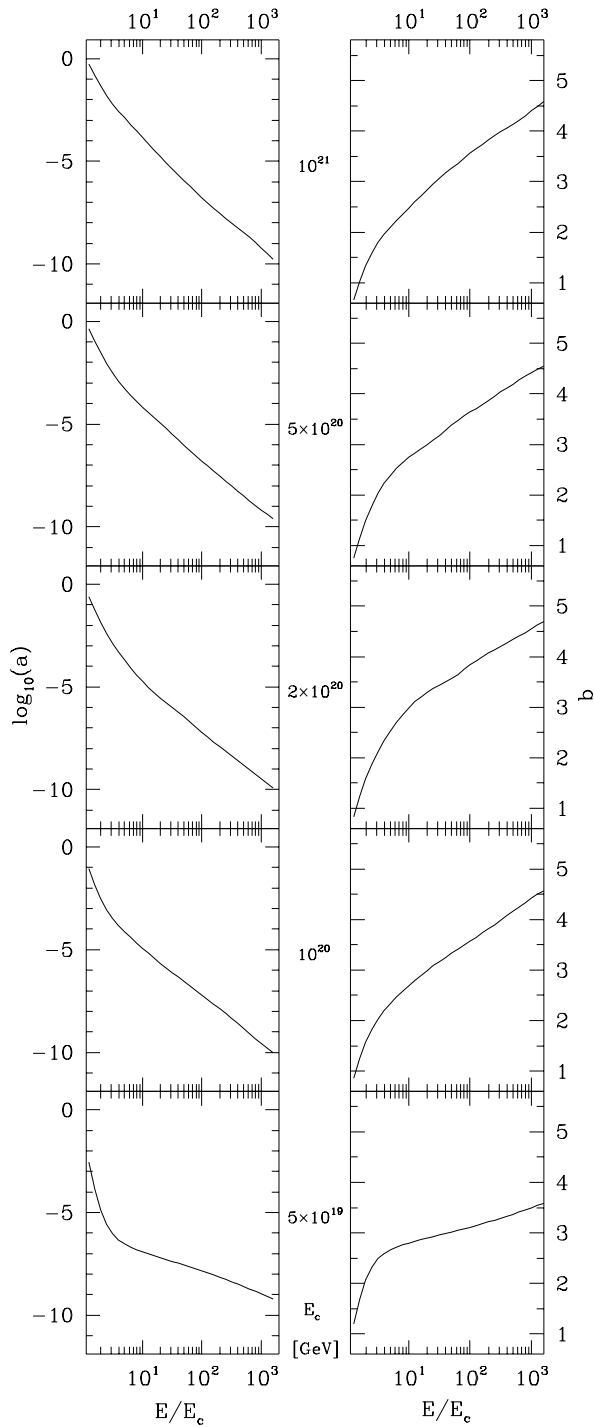,width=8.5cm}
\caption{\label{gzk}
{ \sl The functions $a(E/E_c)$ --left panel-- and $b(E/E_c)$ 
--right panel-- for the probability distribution function 
$P(r,E,E_c)$ using the parametrization 
$\exp[-a\cdot(r/1\ {\rm Mpc})^b]$ for five different threshold
energies ($5\cdot 10^{19}$~eV, $10^{20}$~eV, $2\cdot 10^{20}$~eV,
$5\cdot 10^{20}$~eV and $10^{21}$~eV).
}}
\end{center}\end{figure}

Since it is rather
practical to use the $P(r,E,E_c)$ probability distribution 
function we extended the results of \cite{BW99} by using our Monte-Carlo
technique for UHECR propagation. In order to
cover a much broader energy range than the parametrization
of (\ref{bahcall}) we  used the following type of function
\begin{equation} \label{parametr}
P(r,E,E_c)=\exp\left[ -a\cdot(r/1\ {\rm Mpc})^b\right].
\end{equation}
Fig. \ref{fit} demonstrates the reliability of this parametrization. The
direct Monte-Carlo points and the fitted function (eqn. (\ref{parametr}) with
$a=0.0019$ and $b=1.695$) are plotted for
$E_c=10^{20}$eV and $E=2\cdot 10^{20}$eV.
Fig. \ref{gzk} shows the functions $a(E/E_c)$ and $b(E/E_c)$
for a range of three orders of magnitude and for five different
threshold energies. Just using the functions of $a(E/E_c)$ and 
$b(E/E_c)$, thus a parametrization of $P(r,E,E_c)$ one can obtain the 
observed energy spectrum for any injection spectrum without additional
Monte-Carlo simulation.

\section{RESULTS}\label{sec_res}

\begin{figure}\begin{center}
\epsfig{file=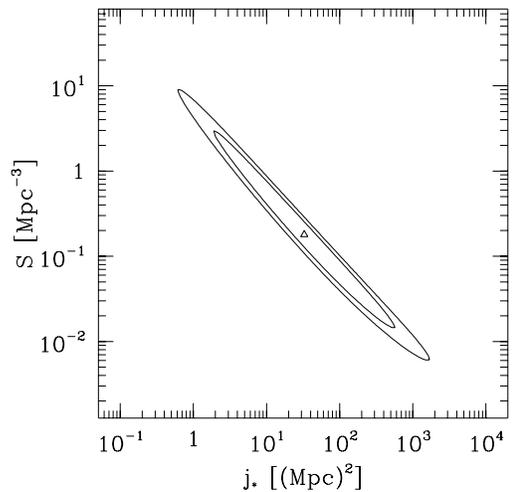,width=7.0cm}
\caption{\label{ell}
{ \sl The $1\sigma$ (68\%) and $2\sigma$ (95\%) confidence
level regions for $j_*$ and the
source density (14 UHECR with one doublet). 
The most probable value is represented by the triangle. The
upper and lower boundaries of these 
regions give for the source density
$180_{- 165( 174)}^{+2730(8817)}\cdot 10^{-3}$~Mpc$^{-3}$
on the 68\% (95\%)confidence
level.
}}
\end{center}\end{figure}

In order to determine the confidence intervals for the  
source densities we used the frequentist method\cite{PDG}.
We wish to set limits
on S, the source density. Using our Monte-Carlo based
$P(r,E,E_c)$ functions and our analytical technique we
determined $p(N_1,N_2,N_3,...;S;j_*)$, which gives the probability of 
observing $N_1$ singlet, $N_2$ doublet, $N_3$ 
triplet etc. events
if the true value of the density is $S$ and the central value of
luminosity is $j_*$. 
The probability distribution is 
not symmetric and far from being Gaussian. For a given set of 
$\{N_i,i=1,2,...\}$ the above probability distribution as a 
function of $S$ and $j_*$ determines the 68\% and 95\%
confidence level regions in the $S-j_*$ plane.
Fig. \ref{ell} shows these regions for our
``favorite'' choice of model ($c(E) \propto E^{-3}$ and
Schechter's luminosity distribution)
and for the present statistics (one doublet out of 14 UHECR events).
The regions are deformed, thin ellipse-like objects in the
$\log(j_*)$ versus $\log(S)$ plane. Since $j_*$
is a completely unknown and independent physical quantity the source density
can be anything between the upper and lower parts of the confidence
level regions. For this model our final answer for the density is
$180_{-165(174)}^{+2730(8817)}\cdot 10^{-3}$~Mpc$^{-3}$,
where the first errors
indicate the 68\%, the second ones in the parenthesis the 95\%
confidence levels, respectively.
The choice of \cite{DTT00} --Dirac-delta like luminosity distribution--
and, for instance, conventional $E^{-2}$ energy distribution
gives much smaller value:
$2.77_{-2.53(2.70)}^{+96.1(916)} 10^{-3}$~Mpc$^{-3}$.
For other choices of $c(E)$
and $h(j)$ see Table \ref{results}. Our results for the Dirac-delta luminosity
distribution are in agreement with
the result of \cite{DTT00} within the error bars. Neverthless, there is a
very important message.
The confidence level intervals are so large, that on the 95\%
confidence level two orders of magnitude smaller densities than
suggested as a lower bound by \cite{DTT00} are also possible.

As it can be seen there is 
a strong correlation between the luminosity and the
source density. Physically it is easy to understand the picture.
For a smaller source density the luminosities should be larger to give the
same number of events. However it is not possible to produce the same
multiplicity structure with arbitrary luminosities.
Very small luminosities can not give multiplets at all,
very large luminosities tend to give more than one doublet.

\begin{figure}\begin{center}
\epsfig{file=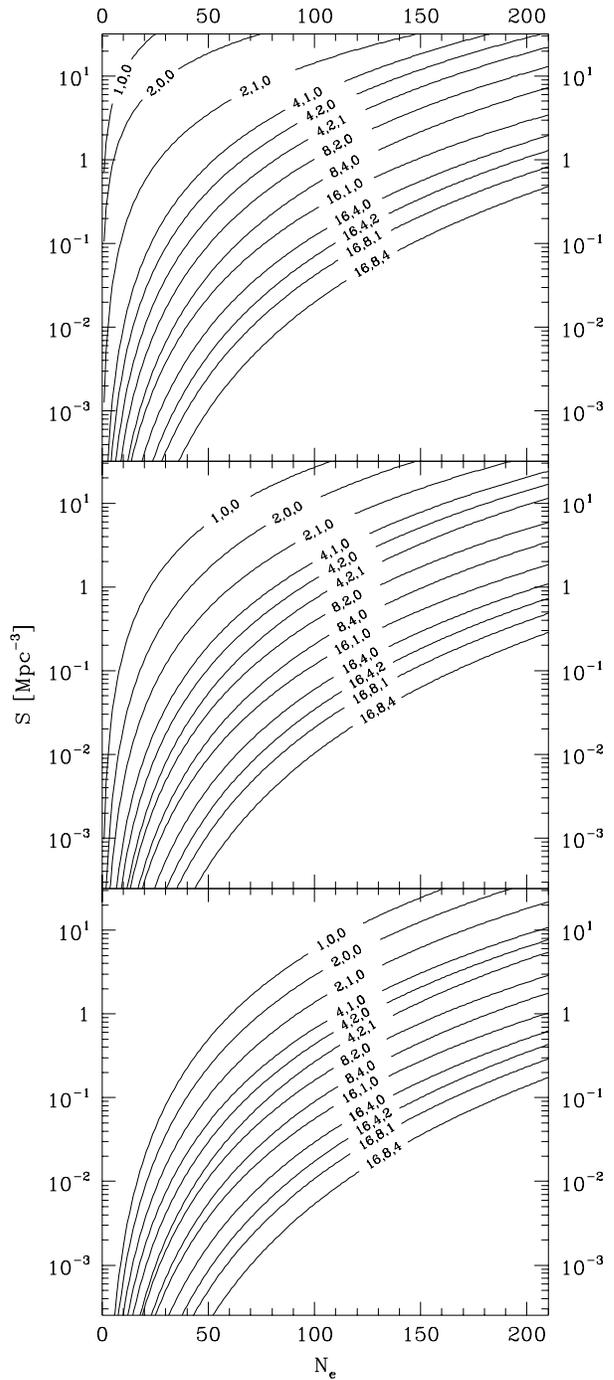,width=8.5cm}
\caption{\label{center}
{ \sl The most probable values for the density of sources 
as a function
of the total number of events (middle panel). The number of multiplets are
indicated on the individual lines in the form: $N_2,N_3,N_4$, where
$N_2,N_3$ and $N_4$ represent the appropriate values for doublets,
triplets and quartets.
The upper and lower panels correspond to the 84 percentile and
16 percentile lines (upper and lower bounds of the 68\% confidence
intervals), respectively.
}}
\end{center}\end{figure}

The same technique can be applied for any hypothetical experimental
result.
For fixed $\{N_k\}$ 
the above probability function determines the 68\% 
confidence regions in $S$ and $j_*$. 
Using these regions one can
tell the 68\% confidence interval for S. The most probable values 
of the source densities for fixed number of multiplets
are plotted on Fig. \ref{center} with the lower and upper bounds. 
The total number of events is shown on the horizontal axis, whereas 
the number of multiplets label the lines. Here again, our ''favorite''
choice of distribution functions were used: $c(E) \propto E^{-3}$ and
$h(j)$ of eqn. (\ref{spread}).

It is of particular interest to analyze in detail the present 
experimental situation having one doublet out of 14 events. 
Since there are some new unpublished events, too, we study the a
hypothetical case of one or two doublets out of 24 events. 
The 68\% and 95\% confidence level results are summarized in Table 
\ref{results} for our three energy and two luminosity distributions. 
It can be seen that Dirac-delta type luminosity distribution
really gives smaller source densities than broad luminosity 
distribution, as it was proven by \cite{DTT00}.
Less pronounced is the effect on the energy distribution of the emitted
UHECRs. The $c(E) \propto E^{-3}$ case gives 
somewhat larger values than the other two choices 
($c(E) \propto E^{-2}$ or given by the decay of a superheavy 
particle). 
The confidence intervals are typically very large, on the 95\%
level they span 4 orders of magnitude. An 
interesting feature of the  results is that ''doubling'' the present
statistics with the same clustering features (in the case studied by 
the table this means one new doublet out of 10 new events) reduces
the confidence level intervals by an order of magnitude. The reduction
is far less significant if we add singlet events only. Inspection of
Fig. \ref{center}  leads to the coclusion that experiments in
the near future
with approximately 200 UHECR events can tell at least the order of 
magnitude of the source density.

\begin{table}[htb]
\begin{center}\begin{tabular}{c|c|c}
$c(E)$ & $h(j)$ & {\bf 14 events 1 doublet} \\
\hline $\propto E^{-2}$ & $\propto \delta$ &
$      2.77 _{-       2.53 (       2.70 )}^{+      96.1  (     916 )}  $\\
\hline $\propto E^{-2}$ & $\propto$ SLF &
$       36.6 _{-        34.3 ( 35.9)}^{+844(4268)}$ \\
\hline $\propto E^{-3}$ & $\propto \delta$ &
$      5.37 _{-       4.98 (       5.25 )}^{+      80.2  (     624 )}  $\\
\hline $\propto E^{-3}$ & $\propto$ SLF &
$       180 _{- 165 ( 174)}^{+2730(8817)}$      \\
\hline $\propto$ decay  & $\propto \delta$ &
$      3.61 _{-       3.30 (       3.51 )}^{+     116    (    1060 )}  $\\
\hline $\propto$ decay  & $\propto$ SLF &
$       40.9 _{-        38.3 ( 40.1)}^{+856(4345)}$ \\
\hline\hline
$c(E)$ & $h(j)$ & {\bf 24 events 1 doublet} \\
\hline $\propto E^{-2}$ & $\propto \delta$ &
$     17.4  _{-      16.0  (      17.0  )}^{+     298    (    2790 )}  $\\
\hline $\propto E^{-2}$ & $\propto$ SLF &
$       200 _{- 169 ( 182)}^{+1230(2428)}$      \\
\hline $\propto E^{-3}$ & $\propto \delta$ &
$     25.0  _{-      22.6  (      24.3  )}^{+     211    (    1690 )}  $\\
\hline $\propto E^{-3}$ & $\propto$ SLF &
$       965 _{- 741 ( 821)}^{+3220(5613)}$      \\
\hline $\propto$ decay  & $\propto \delta$ &
$     20.4  _{-      18.6  (      19.9  )}^{+     358    (    3190 )}  $\\
\hline $\propto$ decay  & $\propto$ SLF &
$       211 _{- 174 ( 190)}^{+1110(2274)}$      \\
\hline\hline
$c(E)$ & $h(j)$ & {\bf 24 events 2 doublets} \\
\hline $\propto E^{-2}$ & $\propto \delta$ &
$      3.19 _{-       2.68 (       2.99 )}^{+      26.4  (     253 )}  $\\
\hline $\propto E^{-2}$ & $\propto$ SLF &
$       41.5 _{-        36.4 ( 40)}^{+424(1514)}$   \\
\hline $\propto E^{-3}$ & $\propto \delta$ &
$      6.42 _{-       5.46 (       6.07 )}^{+      46.2  (     193 )}  $\\
\hline $\propto E^{-3}$ & $\propto$ SLF &
$       208 _{- 182 ( 201)}^{+1970(3858)}$     \\
\hline $\propto$ decay  & $\propto \delta$ &
$      4.18 _{-       3.51 (       3.92 )}^{+      34.5  (     296 )}  $\\
\hline $\propto$ decay  & $\propto$ SLF &
$       45.4 _{-        39.7 ( 43.7)}^{+457(1556)}$ \\
\end{tabular}
\vspace{0.3cm}
\caption{\label{results}
{ \sl The most probable values for the source densities
and their error bars given by the 68\% and 95\% confidence
level regions (the latter in parenthesis). 
The numbers are in units of $10^{-3}$~Mpc$^{-3}$
The three possible energy spectrums 
are given by a distribution proportional to $E^{-2}$, $E^{-3}$, 
or by the decay of a $10^{12}$ GeV particle (denoted by
``decay''). The luminosity distribution can be proportional 
to a Dirac-delta or to Schechter's luminosity function 
(denoted by ``SLF'').
Results are listed for the observed 1 doublet out of 14 events and
for two hypothetical cases (1 doublet out of 24 events and 2 doublets out 
of 24 events). 
}}
\end{center}\end{table}

\section{SUMMARY} \label{sec_sum}

We presented a technique in order to statistically analyze the
clustering features of UHECR. The technique can be applied
for any model of UHECR assuming small deflection. The key role 
of the analysis is played by the $P_k$ functions defined
by eqn. (\ref{P_k}), which is the probability of detecting $k$
events above the threshold from a single source. Using a 
combinatorial expression of eqn. (\ref{distribution}) the 
probability distribution for any set of multiplets can be given
as a function of the source density.

We discussed several types of energy and luminosity distributions
for the sources and gave the most probable source densities
with their confidence intervals for present and future 
experiments. 

The probability $P(r,E,E_c)$ that a proton created at a distance $r$
with energy $E$ arrives above the threshold $E_c$ \cite{BW99} is
determined and parametrized for a wide range of threshold energies. 
This result can be used to obtain the observed energy spectrum of the UHECR
for arbitrary injection spectrum. 

In ref. \cite{DTT00} the authors analyzed the statistical features
of clustering of UHECR, which provided constraints on astrophysical 
models of UHECR when the number of clusters is small, by giving a 
bound from below. In our paper we have shown that there is some 
constraint, but it is far from being tight. At present statistics 
the 95\% confidence level regions usually span 4 orders of magnitude. 
Two orders of magnitude smaller numbers than the prediction
of \cite{DTT00} (their eqn. (13) suggests for the density 
of sources $\sim 6\cdot 10^{-3}$~Mpc$^{-3}$) can also be obtained.  
Adding 10 new events with an additional doublet 
the confidence interval can be reduced to 3 orders of
magnitude and the increase of the UHECR events to 200 can tell
at least the order of magnitude of the source density. 

\section{ACKNOWLEDGEMENTS}

We thank K. Petrovay for clarifying some issues in spherical astronomy.
This work was partially supported by Hungarian Science Foundation
grants No. OTKA-T29803/T22929-FKP-0128/1997.


\begin{thebibliography}{99}
\bibitem{GZK66} K. Greisen, Phys. Rev. Lett. 16 (1966) 748; \\
G.T. Zatsepin and V.A. Kuzmin, Pisma Zh. Exp. Teor. Fiz. 4 (1966) 114.
\bibitem{AGASA} M. Takeda et al., Phys. Rev. Lett. 81 (1998) 1163;\\
astro-ph/9902239; 
www-akeno.icrr.u-\\tokyo.ac.jp/AGASA/results.html\#100EeV
\bibitem{FLY} D.J. Bird et al., Phys. Rev. Lett. 71 (1993) 3401;
Astrophys J. 424 (1994) 491; ibid 441 (1995) 144.
\bibitem{HAVERAH} M.A. Lawrence, R.J.O. Reid and A.A. Watson,
J. Phys. G 17 (1991) 773.
\bibitem{YAKUTSK} N.N. Efimov et al. in Proc. Astrophysical
Aspects of the Most Energetic Cosmic Rays, p. 20, eds. M. Nagano and 
F. Takahara, World Sci., Singapore, 1991.
\bibitem{HIRES} D. Kieda et al., Proc. of the 26th ICRC, Salt Lake,
1999; www.physics.utah.edu/Resrch.html
\bibitem{YT93} S. Yoshida, M. Teshima, Prog. Theor. Phys. 
89 (1993) 833;\\
F.A. Aharonian, J.W. Cronin, Phys. Rev. D50 (1994) 1892;\\
R.J. Protheroe, P. Johnson, Astropart. Phys. 4 (1996) 253.
\bibitem{FP00} G.R. Farrar and T. Piran, Phys. Rev. Lett.
84 (2000) 3527; \\
A. Dar, astro-ph/0006013.
\bibitem{M95} K. Mannheim, Astropart. Phys. 3 (1995) 295.
\bibitem{RB93} J.P. Rachen, P.L. Biermann, Astron. Astrophys. 272
(1993) 161.
\bibitem{ES95} J.W. Elbert, P. Sommers, Astrophys. J. 441 (1995) 151.
\bibitem{B99} R.D. Blandford, Phys. Scr. T85 (2000) 191Phys. Scr. T85 
(2000) 191.
\bibitem{WFP97} E. Waxman, K.B. Fisher and T. Piran, Astrophys. J. 483
(1997) 1; \\
M. Giller, J. Wdowczyk and A. Wolfendale, J. Phys. G6 (1980) 1561; \\
C.T. Hill and D.N. Schramm, Phys. Rev. D31 (1985) 564.
\bibitem{BW99} J.N. Bahcall and E. Waxman, hep-ph/9912326.
\bibitem{Uchi} Y. Uchihori et al., Astropart. Phys. 13 (2000) 151-160.
\bibitem{Hea96} N. Hayashida et al., Phys. Rev. Lett. 77
(1996) 1000.
\bibitem{DTT00} S.L. Dubovsky, P.G. Tinyakov and I.I. Tkachev,
Phys. Rev. Lett. 85 (2000) 1154.
\bibitem{B96} M. Boratav, Nucl. Phys. Proc. 48 (1996) 488.
\bibitem{G99} C.K. Guerard, Nucl. Phys. Proc. 75A (1999) 380.
\bibitem{BBL00} X. Bertou, M. Boratav, A. Letessier-Selvon, \\
Int. J. Mod. Phys. A15 (2000) 2181.
\bibitem{DMS98} B.R. Dawson, R. Meyhandan and K.M. Simpson, Astropart. 
Phys. 9 (1998) 331.
\bibitem{DKDM00} G. Domokos, S. Nussinov, Phys. Lett. B187 (1987) 372;\\
D. Fargion, B. Mele, A. Salis, Astrophys. J. 517 (1999) 725;
T.J. Weiler, Astropart. Phys. 11 (1999) 303, Astropart. Phys. 12 
(2000) 379 (Erratum); \\
G. Domokos, S. Kovesi-Domokos and P.T. Mikulski, hep-ph/0006328.
\bibitem{ELN90} J. Ellis, J.L. Lopez, D.V. Nanopoulos, Phys. Lett.
B247 (1990) 257;\\
J. Ellis et al., Nucl. Phys. B373 (1992) 399;\\
P. Gondolo, G.B. Gelmini, S. Sarkar, Nucl. Phys. B392 (1993) 111.
\bibitem{BKV97} V. Berezinsky, M. Kachelrie{\ss}   and A. Vilenkin, 
Phys. Rev. Lett. 79 (1997) 4302;\\
V.A. Kuzmin, V.A. Rubakov, Phys. Atom. Nucl. 61 (1998) 1028.
\bibitem{BS98} M. Birkel and S. Sarkar, Astropart. Phys. 9 (1998) 297;\\
S. Sarkar, hep-ph/0005256.
\bibitem{BEA}  K. Benakli et al., Phys. Rev. D59 (1999) 047301.
\bibitem{PBRS} P. Blasi and R. Sheth, Phys.Lett. B486 (2000) 233.
\bibitem{Hill} C.T. Hill, Nucl. Phys. B224 (1983) 469.
\bibitem{BK98} V. Berezinsky, M. Kachelriess, Phys. Lett. B434 (1998) 61. 
\bibitem{Schechter} P.L. Schechter, Astrophys. J. 203 (1976) 297.
\bibitem{BS00} P. Bhattacharjee and G. Sigl, Phys. Rep. 327 (2000) 109.
\bibitem{AGNM99} A. Achterberg et al., astro-ph/9907060.
\bibitem{SEMPR00} T. Stanev et al., astro-ph/0003484.
\bibitem{PDG} C. Caso et al. (Particla Data Group), Eur. Phys. J. C3 
(1998) 172.
\end{thebibliography}
\end{document}